\documentstyle[12pt,epsf]{article}
\newcommand {\grav}    {\rm{\tilde G}}

\newcommand {\Zzero}   {{\rm Z}^0}

\newcommand {\ee}         {{e^+e^-}}

\newcommand {\qqb}        {{q\bar{q}}}

\newcommand {\eefg}      {e^+e^-\rightarrow \phi \gamma}

\newcommand {\mydeg}   {^{\circ}}

\begin{document}
\rightline{ LC-TH-2001-015.}
\vspace*{0.2cm}
\rightline{\today}
\vspace*{0.3cm}
\begin{center}
\Huge {\bf Sensitivity to sgoldstino states 
at the future 
linear $\ee$ and photon colliders  
\\}
\vspace{.5 cm}
\large{\bf   Paolo Checchia \\ 
\it I.N.F.N sezione di Padova and Dipartimento di Fisica G.Galilei,\\ Padova, Italy\\ 
\vspace*{0.3cm}

  and\\
  \vspace*{0.3cm}
  
\bf   Enrico Piotto \\
\it CERN, European Organization for Nuclear Research, Geneva, Switzerland.\\

\rm}
\vspace{.5 cm}
\end{center}
\vspace{1. cm}
\begin{abstract}
Sensitivity to the supersymmetric scalar states $\phi$
at the future linear $\ee$ and photon colliders is discussed.
In particular it is illustrated a search strategy  for massive sgoldstinos, the supersymmetric
 partners of the goldstino.

\end{abstract}
\newpage

\section{ Introduction}

In the Supersymmetric extension of the Standard Model, once Supersymmetry is
spontaneously broken the gravitino $\grav$ can acquire a mass absorbing the
degrees of freedom of the goldstino.
The mechanism is analogous to
the spontaneous breaking of the electro-weak 
symmetry in the Standard Model, when Z and W bosons 
acquire mass absorbing the goldstone bosons.

A very light gravitino $\grav$ as predicted by supersymmetric 
models ~\cite{ref:GMSB} has been searched for at LEP and Tevatron 
experiments~\cite{ref:lepbound,ref:cmsbound} and the sensitivity
to its signatures of an experiment at a future linear collider 
has been studied ~\cite{ref:gravmio}.
Limits on the $\grav$ mass are related to 
the supersymmetry-breaking scale $\sqrt{F}$.

It has been pointed out \cite{ref:prz} that in such
supersymmetric extensions  
of the  Standard Model 
with a  light gravitino,
the effective theory at the weak scale
must contain also the supersymmetric partner of the goldstino,
called sgoldstino. The production of this particle, 
which could be massive,
may be relevant at the LEP and Tevatron energies \cite{ref:przhad}
if the supersymmetry-breaking scale and the sgoldstino mass are not too large.
Two states are considered in \cite{ref:prz,ref:przhad}, S CP-even and P
CP-odd. Assuming R-parity conservation, it has to be noticed that, while
the goldstino is R-odd,  the sgoldstino is R-even and therefore 
it can be produced  together with Standard Model particles. 

At  LEP 2 
sgoldstino signatures have been searched for
by  the DELPHI experiment 
\cite{ref:delphisgold}
and preliminary results from CDF
\cite{ref:cdfsgold} show the
higher sensitivity of hadron colliders.
None of the two searches found an evidence for such states.

At an $\ee$ collider
one of the most interesting channels for the production of such  scalars (from now on the symbol
$\phi$ will be used to indicate a generic state) is
the process  $\eefg$ which depends on the $\phi$ mass $m_{\phi}$ and on
$\sqrt{F}$:

\begin{equation}
\frac{d \sigma} {dcos\theta} (e^+e^-\rightarrow \phi \gamma  )
=\frac{\left|\Sigma\right|^2 s}{64 \pi F^2} 
\left( 1- \frac{m_{\phi}^2}{s} \right)^3 (1+cos^2\theta)
\label{dsigma}
\end{equation}
where $\theta$ is the scattering angle in the centre-of-mass and

\begin{equation}
\left|\Sigma\right|^2=\frac{e^2 M_{\gamma\gamma}^2}{2s}+
                      \frac{g_Z^2(v_e^2+a_e^2) M_{\gamma Z}^2 s}{2(s-m_Z^2)^2}+
                      \frac{e g_Z v_e M_{\gamma\gamma}M_{\gamma Z}}{s-m_Z^2}
\end{equation}
with $v_e=sin^2 \theta_W -1/4$, $a_e=1/4$ and 
$g_Z=e/(sin \theta_W cos \theta_W)$.
The parameters $M_{\gamma\gamma}$ and  $M_{\gamma Z}$
are related to the diagonal mass term for the $U(1)_Y$ and $SU(2)_L$ gauginos
$M_1$ and $M_2$:

\begin{equation}
M_{\gamma\gamma}= M_1 cos^2 \theta_W+ M_2 sin^2 \theta_W,~
M_{\gamma Z}= (M_2-M_1) sin \theta_W cos \theta_W.
\end{equation}

Other interesting processes are due to $\gamma \gamma$- or gg-fusion
occurring, respectively, at $\ee$ and hadron colliders.
In both cases the production cross sections are proportional to the corresponding widths:
\begin{equation}
\sigma(\ee \rightarrow \ee  \phi)\propto \sigma_0^{\gamma \gamma}=\frac{4 \pi^2}{ m^3_{\phi}}\Gamma(\phi \rightarrow \gamma \gamma),
\sigma(p\bar{p} \rightarrow \phi) \propto \sigma_0^{g g}=\frac{\pi^2}{8 m^3_{\phi}}\Gamma(\phi \rightarrow g g) 
\end{equation}  
and they can be obtained, respectively, from the photon and gluon distribution functions.

The  decay modes  $\phi \rightarrow \gamma \gamma$ and 
$\phi \rightarrow gg$ widths 
are 
\begin{equation}
\Gamma(\phi\rightarrow \gamma \gamma)=\frac{m_{\phi}^3 M_{\gamma\gamma}^2}{32 \pi F^2}
\label{phitogam}
\end{equation}
and 
\begin{equation}
\Gamma(\phi \rightarrow g g)= \frac{m_{\phi}^3 M_3^2}{4 \pi F^2}
\end{equation}
where $M_3$ is the gluino mass. 

As noticed in \cite{ref:przhad}
the production formulae are  
similar in form to those for a light SM Higgs production in Born approximation where 
$\Gamma(H\rightarrow\gamma \gamma)$ and  $\Gamma(H\rightarrow g g)$
substitute the $\phi$ widths.
It is straightforward to apply the same correspondence between these two different physical 
cases to the $\phi$ production on photon colliders. With a reverse substitution, an
effective  production cross section in the narrow-width approximation can be deduced
from the studies of Higgs Physics at a $\gamma \gamma$ collider~\cite{ref:Telnov}:
\begin{equation}
\sigma^{eff}=\frac{dL_{\gamma \gamma}}{dW_{\gamma \gamma}}\frac{m_{\phi}}{L_{\gamma\gamma}}  \times
             \frac{4\pi^2 \Gamma(\phi\rightarrow \gamma \gamma)}{m_{\phi}^3}
\label{sggphi}
\end{equation}
where $dL_{\gamma \gamma}/dW_{\gamma \gamma}$ is the luminosity spectrum in the 
two photon center-of-mass $W_{\gamma \gamma}$  and $L_{\gamma\gamma}$ 
is defined as the luminosity   
at the high  $\gamma \gamma$ energy peak.

All the above formulae depend on model dependent mass parameters.
In \cite{ref:prz} two sets for these parameters  
are considered to give numerical examples.
They are reported in Table. \,\ref{tab:param}. 

\begin{table}[bth]
\begin{center}
\begin{tabular}{|c|c|c|c|}
\hline
        & $M_1$ & $M_2$        & $M_3$ \\
\hline
1)      & 200   & 300         & 400  \\
\hline
2)      & 350   & 350         & 350  \\ 
\hline
\end{tabular}
\caption[]{Two choices for the gaugino 
mass parameters (in GeV) relevant for the sgoldstino production and decay. }
\label{tab:param}
\end{center}
\end{table}

The total width
for  a large interval of the parameter space is dominated by
$\Gamma(\phi \rightarrow g g)$ and  it is  narrow 
(below the few GeV order) except  for the region with small $\sqrt{F}$
where the production cross section is expected to be very large.


In this note the sensitivity to these states  of an experiment 
at a $\ee$ linear collider with a center-of mass energy of 500 GeV
and the sensitivity of an experiment at a photon collider obtained 
from the same energy primary $\ee$ 
beams are evaluated. 
An integrated  luminosity of 500 fb$^{-1}$ for the $\ee$ collisions
is considered with a reduction factor for the $\gamma \gamma$ interactions.

\section{$\ee$ collider}

The search for these scalars at a future linear collider can be
an upgrade of the analysis done at LEP
where the two decay channels $\phi \rightarrow \gamma \gamma$ 
and $\phi \rightarrow gg$
were considered \cite{ref:delphisgold}.  For the present sensitivity evaluation
only the dominant channel is considered. 
The
 $\phi \rightarrow  g g$ decay gives rise to events with one
photon 
and two jets.
An irreducible background from 
$e^+ e^- \rightarrow q \bar{q} \gamma$ events is associated to this topology
and therefore the signal must be searched for as an excess of events 
over the  background expectations for every mass hypothesis.
To  select    $g g \gamma$ candidate events 
the following selection criteria can be defined:
\begin{itemize}
\item
an electromagnetic energy  cluster identified as photon with 
a polar angle 
 $\theta>20^{\circ}$;
the angle between 
the photon
and the nearest jet must be greater than $10^{\circ}$;
\item  no electromagnetic cluster with $\theta< 5^{\circ}$;
\item to remove $\gamma \gamma$ fusion  events:
the total multiplicity  $>10$;
the charged multiplicity  $> 5$;
the energy in transverse plane $> 0.12\cdot \sqrt{s}$;
the sum of absolute values of track momentum 
along thrust axis $>0.20 \cdot\sqrt{s}$;
\item to  remove  Bhabha background:
reject the events with electromagnetic cluster with $E> 0.45 \cdot \sqrt{s}$ and
low track multiplicity;
\item to reduce $\qqb \gamma$ events:  $ |cos(\theta_p)|<.995 $ where $\theta_p$ is the polar angle of
missing momentum;
the visible energy  greater than $0.60 \cdot \sqrt{s}$;
reject events with c or b tag;
\item to remove WW background
the events are reconstructed forcing into 2 jets topology 
      but removing from jetization the tracks associated to the photon
      cluster. Events are removed if $ y_{cut}>0.02$.


\end{itemize}

The  polar angle acceptance for a $\phi \gamma$ signal produced as in (\ref{dsigma})
is about $80 \%$. It has been evaluated by generating 4-vectors corresponding 
to the prompt photon and to the $\phi$ decay products.  
Considering the DELPHI results \cite{ref:delphisgold}, the selection efficiency inside
the acceptance region 
is assumed to be
of the order of 50 $\%$.

\begin{figure}[th]
\begin{center}\mbox{\epsfxsize 9cm\epsfbox{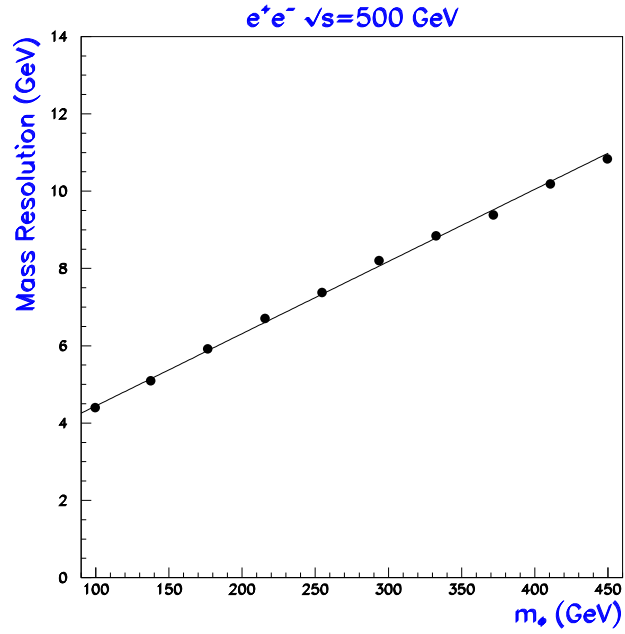}} \end{center}
\caption{
Mass resolution as function of the considered $\phi$ mass hypothesis. 
The full line corresponds to a linear fit.}
\label{m_res}
\end{figure}

The associated photon is monochromatic (except for the region  with small $\sqrt{F}$
where the production cross-section is expected to be very large) 
for a given center-of-mass energy. 
Therefore the signal  can be detected as a peak in the photon energy distribution 
of the selected events. In addition, the photon energy could be determined very precisely
by means of kinematic constraints if a final state three body topology is assumed.
However, the presence of the beamstrahlung ($2.8 \%$ of mean beam-energy 
loss~\cite{ref:brink}) induces a smearing on the photon energy 
which is comparable with or larger than
the experimental resolution.
On the other hand, the signal can be searched for directly in the jet-jet  
invariant mass distribution. Clearly the detector performance plays a crucial role
in the optimal search strategy. Here a jet energy resolution following the
$\sigma_E^{jet}/E=40\%/\sqrt{E} \oplus 2\%$ dependence and an error of about one degree
in the jet angle reconstruction is assumed. With these assumptions
the direct mass search is convenient or comparable w.r.t. the
recoil photon search. The mass resolution is given in Fig. \ref{m_res}.
%

\begin{figure}[th]
\begin{center}\mbox{\epsfxsize 9cm\epsfbox{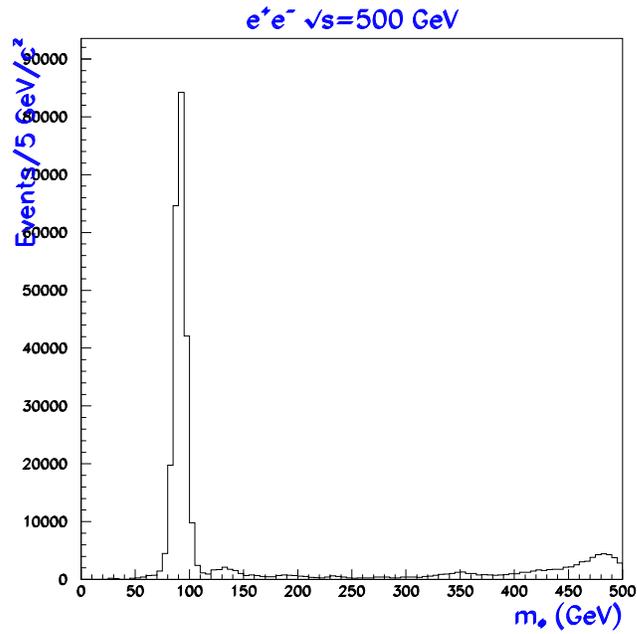}} \end{center}
\caption{
 Jet-jet invariant mass spectrum for the 
$\qqb \gamma $ events.   }
\label{ms_lc}
\end{figure}

The  background rate depends 
on the considered $\phi$ mass hypothesis as it  can be seen in Fig. \ref{ms_lc} where
the reconstructed jet-jet invariant mass of 
 $\qqb \gamma$ events generated with PYTHIA \cite{ref:pythia} in the acceptance region is shown.
The  events are scaled in order to reproduce the number of expected events with an integrated
$L_{\ee}$ luminosity of 500 fb$^{-1}$. However the statistical fluctuations are not reproduced.   

Given the background event distribution as function of $m_{\phi}$ and  
the detection efficiency for any  $\phi$ mass hypothesis it is possible to estimate 
a $95\%$ Confidence Level  cross section limit for the  $\phi$ production cross section.
Only statistical fluctuations are considered here. The bin to bin 
fluctuations on the number of background events due to the reduced Monte Carlo statistics
are removed by a spline function. 

\begin{figure}[th]
\begin{center}\mbox{\epsfxsize 8cm\epsfbox{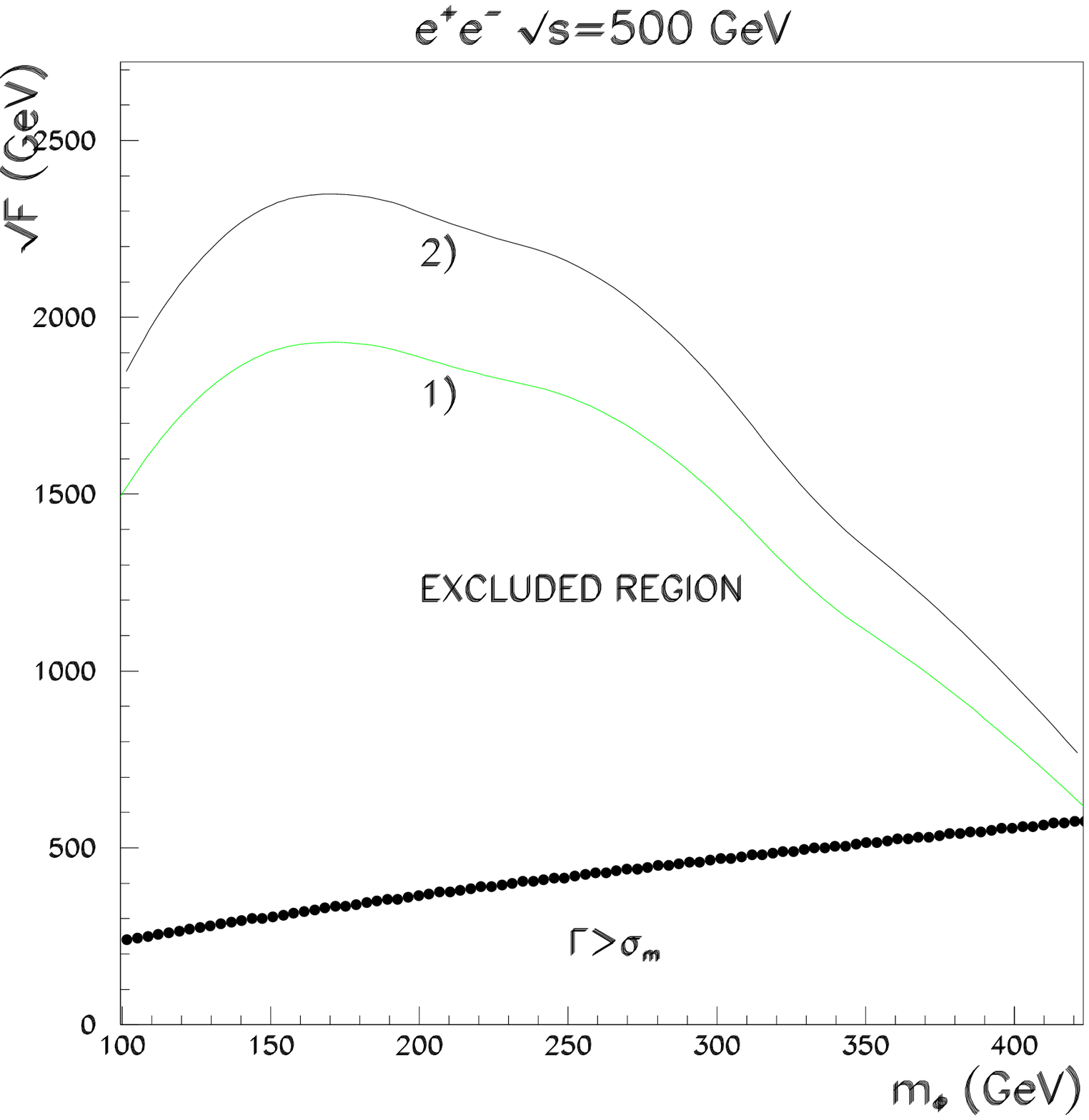}\epsfxsize 8cm\epsfbox{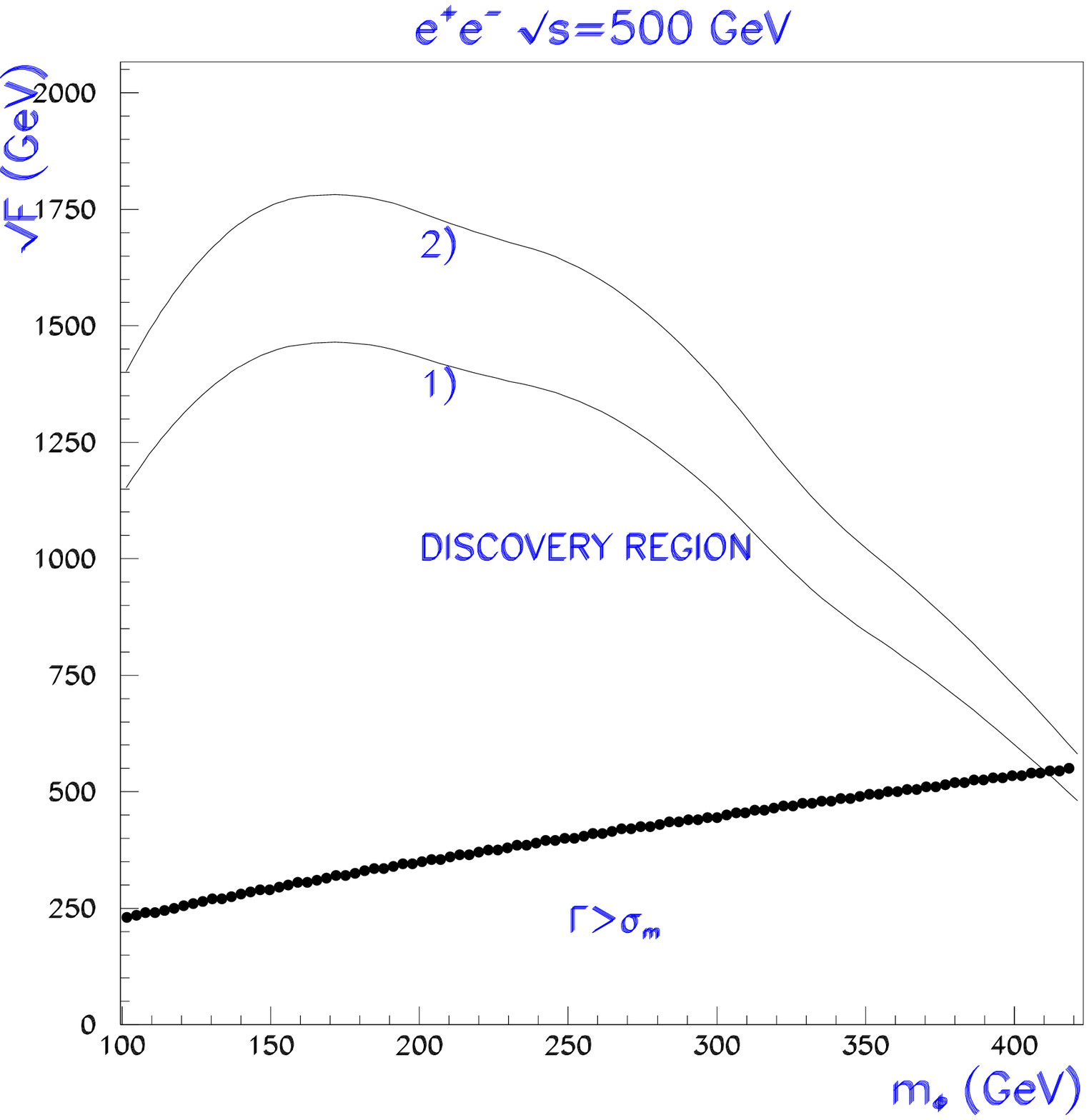}} \end{center}
\caption{
 Exclusion region at the 95$\%$ Confidence Level
and $>5~\sigma$ signal discovery region 
 in  $m_{\phi}$  $\sqrt{F}$ space 
 for the two
 sets of parameters of Tab. \ref{tab:param}. The thick lines indicate the region
 where the decay width $\Gamma$ is larger than the experimental resolution.  
}
\label{excl}
\end{figure}

By comparing the experimental limits with the production cross section 
computed from (\ref{dsigma}) 
it is possible to determine a  $95 \%$ Confidence Level
excluded region on the parameter space and a $5~\sigma$ discovery region. 
The beamstrahlung effects which are more relevant than the 
Initial State Radiation one's
are taken into account.
The limit  and the $5~\sigma$ regions are shown in Fig. \ref{excl}. 
The $\phi$ width for all the considered  $m_{\phi}$ values is 
smaller than the 
experimental resolution in all the points corresponding to the   
limit curves. Therefore 
the limit has been computed integrating the signal only over the
experimental resolution.
The region where the expected width is larger than the experimental resolution
is indicated in  Fig. \ref{excl}.
For $m_{\phi}<420$ it is
 possible to cover this region of parameter space given the high cross section.
This is no more true for $m_{\phi}> 420$ GeV where the decreasing cross section   and
the increasing  width   result in a drop of experimental sensitivity. 



In the near future the Fermilab Tevatron Collider is expected to increase the luminosity 
by a factor $\sim 20$ \cite{ref:tdrcdf} and consequently an increase of  about 1.5 in their $\sqrt{F}$ 
limits can be envisaged.
The limits shown in Fig. \ref{excl} are then competitive with the future 
improved Tevatron results.

At $\ee$ colliders,
additional information can be obtained by  
searching for  the associated $\phi \Zzero$ production as described in \cite{ref:prz}. 
As far as the  production cross section is considered, competitive results are expected
in the $m_{\phi}<\sqrt{s}-m_Z$ region. However, since this channel has  a
different final state topology requiring a more sophisticated analysis, it is  
not considered here.

\section{$\gamma \gamma$ collider}
The effective  cross section given in eq. (\ref{sggphi}) depends on the luminosity 
factor  
$f_L=\frac{dL_{\gamma \gamma}}{dW_{\gamma \gamma}}\frac{m_{\phi}}{L_{\gamma \gamma }} $.
In the photon collider projects \cite{ref:gamgampro} there are several possible scenarios 
concerning the photon energy spectra. It may be desirable a photon energy 
distribution peaked as much as possible toward the primary electron/positron  energy.
In \cite{ref:Telnov}  $f_L=7$ is assumed and $L_{\gamma \gamma}$ is taken as the integral luminosity
for $z>z_{min}=0.65$ where $z=W_{\gamma \gamma}/2E_e$ and $E_e$ is the primary electron beam energy. 
The luminosity high energy peak 
is expected to have a FWHM of 
$\sim 10-15 \%$ with a sharp edge at $z\sim0.8$.
Therefore the unexcluded $m_{\phi}-\sqrt{F}$  parameter space achievable
at these machines with $2E_e=500$ GeV ensures that the $\phi$ width is negligible.

The effective cross section obtained with $f_L=7$ is much higher 
(several orders of magnitude, depending on $m_{\phi}$) than the  
$\ee \rightarrow \phi \gamma$ cross section with the same parameters. 
Considering the photon and gluon decay channels, 
the signal would appear as a peak of two high energy photons 
or  jets with no transversal missing energy. The two jets 
final state has to compete 
with large Standard Model background which can be suppressed using polarized
photon beams  with polarizations $\lambda_1, \lambda_2$:
$\sigma({\gamma\gamma \rightarrow \qqb}) \propto 1-\lambda_1 \lambda_2$
while $\sigma({\gamma\gamma \rightarrow \phi}) \propto 1+\lambda_1 \lambda_2$.  
However, taking into account QCD corrections \cite{ref:bord,ref:jikia2}, the $\qqb g$
final state with unresolved gluon jet gives rise to 
a   sizeable background which may be   hard to reject. 
Therefore, despite of the smaller decay branching ratio,   
only the two photons final state which has a very little Standard Model 
background is considered here.

The selection of events with two collinear high energetic photons
is rather simple and the LEP experience can be used \cite{ref:lepgg}.
An efficient way to select photons and to reject electrons is to require two 
energy clusters in the electromagnetic calorimeter not associated to hits in the 
vertex detector. Events with tracks detected in the other tracking devices only in one 
hemisphere  can be accepted to recover photon conversions.
Other requirements are:
\begin{itemize}
\item acollinearity  between the e.m. clusters smaller than $30 ^{\circ}$;  
\item acoplanarity
smaller than $5^{\circ}$; 
\item polar angle $\theta>30^{\circ}$; 
\item $E_{\gamma}>0.9 \cdot z_{min} \cdot E_e$.
\end{itemize}

The detection efficiency is very high ($> 90\%$) in the region $W_{\gamma \gamma}/2E_e> z_{min}$
and the acceptance for the decay of a scalar particle is $86\%$.
 
The irreducible Standard Model background of $\gamma \gamma \rightarrow \gamma \gamma$
events has been discussed in \cite{ref:Jikia,ref:Gounaris}.    
In the $W_{\gamma \gamma}$ region above   200-250 GeV 
the cross section is in the range  8-14 fb for $\theta > 30 \mydeg$
and then,  
assuming $L_{\gamma \gamma}\sim0.15 \cdot L_{\ee}$, the number of expected events 
is of the order of 600-1000. As a consequence any New Physics signal has
to exceed the corresponding statistical error ( which is of the order of 3 to 4 $\%$) and
the systematic uncertitude including the precision on the background calculation.
For the present sensitivity study an overall background uncertitude of 5 $\%$ is assumed, 
leaving more detailed analysis of the   
signal and background including the comparison of 
their angular distributions to a later stage. 
With these assumptions, the sensitivity to a scalar state decaying in two photons is given
by the expected 95 $\%$  Confidence Level limit on the cross section times 
branching ratio and it is 
\[
\sigma ({\gamma \gamma \rightarrow \phi}) 
\times B.R.(\phi \rightarrow \gamma \gamma) < 1 ~\rm{fb}
\] 
at the 95 $\%$ Confidence Level 
for $m_{\phi} \sim 400$ GeV.

\begin{figure}[th]
\begin{center}\mbox{\epsfxsize 9cm\epsfbox{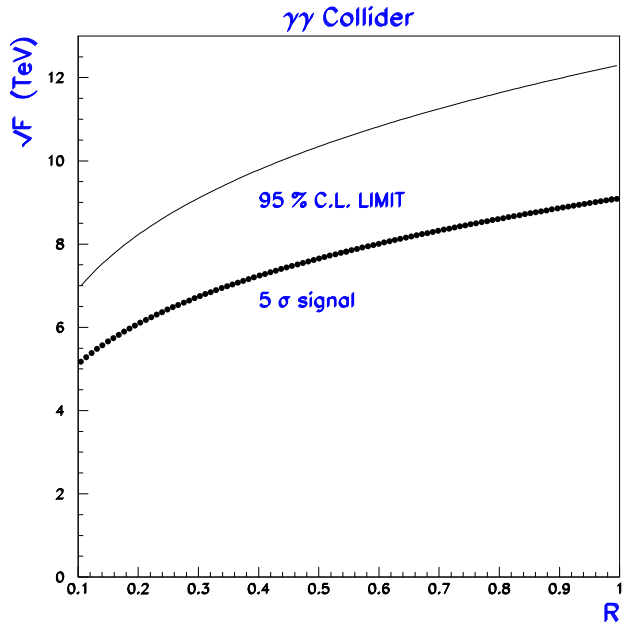}} \end{center}
\caption{
 Limit at 95 $\%$ Confidence Level 
 on the supersymmetry breaking scale and $5~\sigma$ signal   (thick line)
 for the production of a $\sim 400$ GeV
$\phi$ scalar state as function of the ratio $R$.   }
\label{lim_gamgam}
\end{figure}

This value is obtained following the hypothesis that
the whole luminosity is collected at the maximal energy spectrum
available with
250 GeV electron beams. The actual sensitivity for 
several $\phi$ mass hypothesis depends on the machine run strategy, on the available
energy spectrum and on the photon beam polarization. 
Nevertheless, taking the given limit just
as an evaluation of the order of magnitude for the sensitivity, it is 
worth   investigating the 
effect on the supersymmetry breaking scale from 
 (\ref{phitogam}) and (\ref{sggphi}). In particular, defining as  a reference 
cross-section-branching-ratio-product the value $\sigma B$ obtained
with $M_{\gamma \gamma} =350$ GeV
and a 10$\%$ branching ratio to two photons, the limit on
$\sqrt{F}$ and the $5~\sigma$ signal can be expressed in terms of the ratio 
$R=\sigma^{eff} \times B.R.(\phi \rightarrow \gamma \gamma)/\sigma B$.
They are
 then proportional to $R^{\frac{1}{4}}$  
as shown in Fig. \ref{lim_gamgam}.
The sensitivity is clearly much larger than the one expected at the $\ee$ machines.



\section{Conclusions}
The sensitivity to the supersymmetric scalar $\phi$ at the future 
linear $\ee$ and photon colliders is such that unexplored parameter space regions
can be investigated. The $\ee$ machines with center of mass energy of  500 GeV
can set limits for the production of sgoldstino scalars up to about 420 GeV. 
These limits are competitive w.r.t. the expected future results from the Tevatron RUN II.
The sensitivity at the photon colliders obtained from the same electron-positron beam energy
is expected to be much higher for 
$m_{\phi}\sim 400$ GeV. 
 

\subsection*{Acknowledgements}
\vskip 3 mm
We want to thank  F. Zwirner for useful explanations on the
theoretical framework and 
for suggestions on the experimental possibilities, A. Castro for discussions 
concerning the future CDF results and M. Mazzucato for comments
and for reading the manuscript.

\end{document}